\documentclass[twocolumn,nofootinbib,amsmath,prd,aps,superscriptaddress,tightenlines,preprintnumbers]{revtex4}

\pdfoutput=1
\usepackage{graphicx}
\usepackage{amsmath}
\usepackage{amssymb}

\def\bea{\begin{eqnarray}}
\def\eea{\end{eqnarray}}
\def\ba{\begin{eqnarray}}
\def\ea{\end{eqnarray}}
\def\be{\begin{equation}}
\def\ee{\end{equation}}
\def\beq{\begin{equation}}
\def\eeq{\end{equation}}

\newcommand{\lsim}{\mathrel{\rlap{\lower4pt\hbox{\hskip1pt$\sim$}}
    \raise1pt\hbox{$<$}}}         
\newcommand{\gsim}{\mathrel{\rlap{\lower4pt\hbox{\hskip1pt$\sim$}}
    \raise1pt\hbox{$>$}}}         

\newcommand{\leftrightarrowraised}{\mathrel{\rlap{\lower-0pt\hbox{\hskip1pt$\partial$}}
    \raise6 pt\hbox{$\leftrightarrow$}}}

\newcommand{\vect}[1]{\boldsymbol{\rm #1}}

\begin{document}

\newcount\hour \newcount\minute
\hour=\time \divide \hour by 60
\minute=\time
\count99=\hour \multiply \count99 by -60 \advance \minute by \count99
\newcommand{\mydate}{\ \today \ - \number\hour :00}


\title{Astrophysics independent bounds on the annual modulation of dark matter signals}

\def\ific{Dep.\ Fisica Teorica, IFIC, U.\ de Valencia-CSIC, 
Paterna, Apt.\ 22085, 46071 Valencia, Spain}
\def\mpi{Max-Planck-Institut f{\"u}r Kernphysik, Saupfercheckweg 1, 69117 Heidelberg, Germany}
\author{\textbf{Juan Herrero-Garcia}\vspace*{0mm}}
\email{juan.a.herrero AT uv.es}
\affiliation{\ific}

\author{\textbf{Thomas Schwetz}\vspace*{0mm}}
\email{schwetz AT mpi-hd.mpg.de}
\affiliation{\mpi}

\def\Cincy{Department of Physics, University of Cincinnati, Cincinnati, Ohio 45221,USA}
\author{\textbf{Jure Zupan}
\vspace*{3mm}}
\email{zupanje AT ucmail.uc.edu}
\affiliation{\Cincy}

\begin{abstract}
We show how constraints on the time integrated event rate from a given
dark matter (DM) direct detection experiment can be used to set a
stringent constraint on the amplitude of the annual modulation signal
in another experiment. The method requires only very mild assumptions
about the properties of the local DM distribution: that it is
temporally stable on the scale of months and spatially homogeneous
on the ecliptic. We apply the method to the annual modulation
signal in DAMA/LIBRA, which we compare to the bounds derived from the
constraints on the time-averaged rates from XENON10, XENON100, CDMS
and SIMPLE. Assuming a DM mass of 10 GeV, we show that a DM
interpretation of the DAMA/LIBRA signal is excluded at $6.3\sigma$
($4.6\sigma$) for isospin conserving (violating) spin-independent
interactions, and at $4.9\sigma$ for spin-dependent
interactions on protons.

\end{abstract}

\maketitle

{\bf Introduction.} Dark matter (DM) constitutes a significant
fraction of the energy density in the universe, $\Omega_{\rm DM}=0.229
\pm 0.015$ \cite{Komatsu:2010fb}. This conclusion is based entirely on
gravitational effects of DM. A fundamental question is whether DM
interacts also non-gravitationally. There are a number of experiments
searching for signs of such DM interactions.  Direct detection
experiments, for instance, are looking for a signal of DM particles
from the galactic halo that would scatter in underground detectors. A
characteristic feature of the resulting signal will be an annual
modulation, because the earth rotates around the sun, while at the
same time the sun moves relative to the DM halo~\cite{Drukier:1986tm}.
  
At present two experiments are reporting annually modulated signals,
DAMA/LIBRA~\cite{Bernabei:2008yi} (DAMA for short) and
CoGeNT~\cite{Aalseth:2011wp}, with significances of 8.9$\sigma$ and
$2.8\sigma$, respectively. Are these signals due to DM? The answer is
readily obtained by (i) assuming a specific local DM velocity
distribution and (ii) postulating the predominant DM--nucleus
interaction.  Usually a simple Maxwellian DM halo is adopted.  If
interpreted in terms of elastic spin-independent DM scattering both
claims are in tension \cite{Schwetz:2011xm, Fox:2011px} with bounds on
time integrated rates from other direct detection experiments such as
XENON10~\cite{Angle:2011th}, XENON100~\cite{Aprile:2011hi}, or
CDMS~\cite{Ahmed:2010wy}.  The situation may change in the case of
non-standard DM halos with, e.g., highly anisotropic velocity
distributions, DM streams or DM debri flows. Recently CDMS provided a
direct bound on the modulation signal, which disfavors the CoGeNT
modulation without referring to any halo or particle physics model
\cite{Ahmed:2012vq}. Therefore we focus below mainly on DAMA.

In this Letter we present a general method that avoids astrophysical uncertainties
when comparing putative DM modulation signals with the bounds on time
averaged DM scattering rates from different experiments. The method combines 
the results from \cite{Fox:2010bz, Fox:2010bu} with the bounds on the
modulation derived by us in \cite{HerreroGarcia:2011aa}. We are then
able to translate the bound on the DM scattering rate in one
experiment into a bound on the annual modulation amplitude in a
different experiment. The resulting bounds present roughly an order of
magnitude improvement over \cite{Fox:2010bz, Fox:2010bu} and
\cite{HerreroGarcia:2011aa}.

The bounds are (almost completely) astrophysics independent. Only very
mild assumptions about DM halo properties are used: (i) that it does
not change on the time-scales of months, (ii) that the density of DM
in the halo is constant on the scales of the earth-sun distance, and
(iii) that the DM velocity distribution is smooth on the scale of the
earth velocity $v_e= 29.8$~km/s. If the modulation signal is due to
DM, then the modulation amplitude has to obey the bounds. In the
derivation an expansion in $v_e$ over the typical DM velocity $\sim
200$ km/s is used. The validity of the expansion can be checked
experimentally, by searching for the presence of higher harmonics in
the time-stamped DM scattering data \cite{HerreroGarcia:2011aa}.

{\bf Bounds on the annual modulation.} We focus on the case of DM $\chi$
elastically scattering off a nucleus $(A,Z)$ and depositing the nuclear
recoil energy $E_{nr}$ in the detector. The differential rate in
events/keV/kg/day is then given by 
\beq\label{eq:R} 
R_A(E_{nr}, t) =
\frac{\rho_\chi \sigma_A^0}{2 m_\chi \mu_{\chi A}^2} \, F_A^2(E_{nr}) \,
\eta(v_m, t) \,, 
\eeq 
with $\rho_\chi$ the local DM density, $\sigma_A^0$ the total DM--nucleus
scattering cross section at zero momentum transfer, $m_\chi$ the DM mass,
and $\mu_{\chi A}$ the reduced mass of the DM--nucleus system. $F_A(E_{nr})$
is a nuclear form factor. 
For SI interactions with a nucleus $(A,Z)$, $\sigma_A^0$  can be written as
%
$
\sigma^{\rm SI}_A = \sigma_p [Z+(A-Z)(f_n/f_p)]^2 
\mu_{\chi A}^2 / \mu_{\chi p}^2,
$
%
where $\sigma_p$ is the DM--proton cross-section and $f_{n,p}$ are
coupling strengths to neutron and proton, respectively. Apart from a
common overall factor $\rho_{\chi}$ the astrophysics enters the
predicted rate in Eq.~\eqref{eq:R} through the halo integral
\beq\label{eq:eta} 
\eta(v_m, t) \equiv 
\int_{v > v_m} \negthickspace \negthickspace d^3 v 
\frac{f_{\rm det}(\vect{v}, t)}{v} ,
\quad
v_{m}=\sqrt{ \frac{m_A E_{nr}}{2 \mu_{\chi A}^2}},
\eeq
where $v_m$ is the minimal velocity required to have at least $E_{nr}$
energy deposited in the detector. The function $f_{\rm det}(\vect v, t)$
describes the distribution of DM particle velocities in the detector rest
frame with $f_{\rm det}(\vect v, t) \ge 0$ and $\int d^3 v f_{\rm det}(\vect
v, t) = 1$. It is related to the velocity distribution in the rest frame of
the sun by $ f_{\rm det}(\vect{v},t) = f_{\rm sun}(\vect{v} + \vect{v}_e(t))$, where $\vect{v}_e(t)$ is the velocity vector of the earth.  The
rotation of the earth around the sun introduces a time dependence in the
DM-nucleus scattering rate through
$
\eta(v_m,t)=\overline \eta (v_m) + \delta\eta(v_m, t),
$ 
where 
\beq
\delta \eta(v_m, t) = A_\eta(v_m) \cos 2\pi[t - t_0(E_{nr})] ,
\eeq
when expanding to first order in $v_{\rm e}=29.8$ km/s $\ll v_{\rm
sun}\simeq$ 230 km/s. Here, $A_\eta(v_m)$ is defined to be positive.

Let us now assume that $f_{\rm sun}(v)$ is smooth on the scale of
$v_e$, and the only time dependence comes from the rotation of the
earth around the sun and $f_{\rm sun}(v)$ itself is constant in time
and space.
Then the modulation amplitude $A_\eta(v_m)$ can be
bounded in terms of the unmodulated halo integral $\overline\eta$ in
the following way \cite{HerreroGarcia:2011aa}:
\beq \label{eq:bound_gen}
A_\eta(v_m) \leq v_e 
\left[-\frac{d \overline \eta}{d v_m}+ 
    \frac{\overline{\eta}(v_m)}{v_m}  
   - \int_{v_m} dv \frac{\overline{\eta}(v)}{v^2} 
\right] \,.
\eeq
The first term in \eqref{eq:bound_gen} is positive since $\overline
\eta(v_m)$ is a monotonously decreasing function of $v_m$. 
If we further assume that the DM halo is symmetric, so that there is
only one single direction related to the DM flow (see
\cite{HerreroGarcia:2011aa} for details), then one obtaines a more
stringent constraint:
\beq\label{eq:bound_spec}
\int_{v_{1}}^{v_{2}} \negthickspace \negthickspace dv_m A_\eta(v_m)
\leq  \sin\alpha \, v_e\left[\overline \eta(v_{1}) - 
v_{1} \negthickspace \int_{v_{1}} \negthickspace \negthickspace
\negthickspace dv \frac{\overline\eta(v)}{v^2} \right] \,.
\eeq
Here $\alpha$ is the angle between the DM flow and the direction orthogonal
to the ecliptic. The most conservative bound is obtained for $\sin\alpha =
1$ (which would correspond to a DM stream parallel to the ecliptic). However, in many cases the DM
flow will be aligned with the motion of the sun within the galaxy. 
This holds for any isotropic velocity distribution and, up to a
small correction due to the peculiar velocity of the sun, also for tri-axial
halos or a significant contribution from a possible dark-disc. In this case
we have $\sin\alpha \simeq 0.5$. 

In the following we will use time averaged rates from various
experiments to derive an upper bound on $\overline \eta(v_m)$. In
order to be able to apply this information we integrate
Eq.~\eqref{eq:bound_gen} over $v_m$ and drop the negative terms in
Eqs.~\eqref{eq:bound_gen} and \eqref{eq:bound_spec}. 
This gives the bounds
\begin{align}
\int_{v_{1}}^{v_{2}} \negthickspace \negthickspace dv_m A_\eta(v_m) &
\leq  v_e\left[\overline \eta(v_{1}) +
\int_{v_{1}}^{v_{2}} \negthickspace \negthickspace
dv \frac{\overline\eta(v)}{v} \right] \,, 
\label{eq:bound_gen2}\\
\int_{v_{1}}^{v_{2}} \negthickspace \negthickspace dv_m A_\eta(v_m)  &
\leq  \sin\alpha \, v_e \, \overline \eta(v_{1})  \,,
\label{eq:bound_spec2}
\end{align}
In practice the integrals on the l.h.s.\ are replaced by a sum over bins.
Below we will refer to the relations \eqref{eq:bound_gen2} and
\eqref{eq:bound_spec2} as the bounds from  ``general halo'' and ``symmetric
halo'' (where we will take $\sin\alpha = 0.5$), respectively.

{\bf Bounds on the unmodulated halo integral.}  Let us first consider 
SI scattering with $f_n = f_p$. Generalization to isospin violating scattering
with $f_n \neq f_p$ and to SD scattering is straightforward.  The
predicted number of events in an interval of observed energies
$[E_1,E_2]$ is given by
\beq\label{eq:events-pred}
N_{[E_1,E_2]}^{\rm pred}= M T A^2  
\int_0^{\infty} 
\negthickspace\negthickspace\negthickspace dE_{nr} F_A^2(E_{nr})
G_{[E_1,E_2]}(E_{nr}) \tilde\eta(v_m).  
\eeq 
Here $G_{[E_1,E_2]}(E_{nr})$ is the detector response function, which
describes the contribution of 
events with true nuclear-recoil energy $E_{nr}$ to
the observed energy interval $[E_1,E_2]$. It may be
non-zero outside the $E_{nr}\in [E_1,E_2]$ interval due to the finite
energy resolution and includes also (possibly energy dependent)
efficiencies. $M$ and $T$ are the detector mass and exposure time,
respectively, and we defined
\beq\label{eq:etatilde}
\tilde\eta \equiv \frac{\sigma_p \rho_\chi}{2m_\chi \mu^2_{\chi p}} \overline\eta \,,
\eeq
where $\tilde\eta$ has units of events/kg/day/keV. 

Now we can use the fact that $\tilde \eta$ is a falling function
\cite{Fox:2010bz} (see also \cite{Frandsen:2011gi,
  Gondolo:2012rs}). Among all possible forms for $\tilde \eta$ such
that they pass through $\tilde \eta(v_m)$ at $v_m$, the minimal number
of events is obtained for $\tilde \eta$ constant and equal to $\tilde
\eta(v_m)$ until $v_m$ and zero afterwards. Therefore, for a given
$v_m$ we have a lower bound $N_{[E_1,E_2]}^{\rm pred}(v_m) \ge
\mu(v_m)$ with
\beq\label{eq:mu}
\mu(v_m) = M T A^2  \tilde\eta (v_m) \int_0^{E(v_m)} 
\negthickspace dE_{nr} F_A^2(E_{nr}) G_{[E_1,E_2]}(E_{nr}), 
\eeq
where $E(v_m)$ is given in \eqref{eq:eta}. Suppose an experiment
observes $N_{[E_1,E_2]}^{\rm obs}$ events in the intervall $[E_1,E_2]$. Then
we can obtain an upper bound on $\tilde\eta$ for a fixed $v_m$ at a
confidence level CL by requiring that the probability of obtaining
$N_{[E_1,E_2]}^{\rm obs}$ events or less for a Poisson mean of
$\mu(v_m)$ is equal to $1-$CL. Note that this is actually a lower
bound on the CL, since Eq.~\eqref{eq:mu} provides only a lower bound
on the true Poisson mean. For the same reason we cannot use the
commonly applied maximum-gap method to derive a bound on $\tilde\eta$.
If several different nuclei are present, there will be a corresponding
sum in Eqs.~\eqref{eq:events-pred} and \eqref{eq:mu}.

The limit on $\tilde\eta$ can then be used in the r.h.s.\ of
Eq.~\eqref{eq:bound_gen2} or \eqref{eq:bound_spec2} to constrain the
modulation amplitude. For concreteness we first focus on the annual
modulation in DAMA.  If $m_\chi$ is around 10~GeV, then DM particles do not
have enough energy to produce iodine recoils above the DAMA
threshold. We can thus assume that the DAMA signal is entirely due to
the scattering on sodium. We define $\tilde A_\eta \equiv \sigma_p
\rho_\chi /(2m_\chi \mu^2_p) A_\eta$, which is related to the observed
modulation amplitude $A_i^{\rm obs}$ by
\beq\label{eq:Atilde}
\tilde A_\eta^{\rm obs}(v_m^i) = \frac{A_i^{\rm obs} q_{\rm Na}}
{A^2_{\rm Na} \langle F^2_{\rm Na} \rangle_i f_{\rm Na}} \,.
\eeq
Here $q_{\rm Na} = dE_{ee}/dE_{nr}$ is the sodium
quenching factor translating keVee into keVnr, for which we take $q_{\rm Na} =
0.3$. The index $i$ labels energy bins, with $v_m^i$ given by the corresponding energy bin
center using Eq.~\eqref{eq:eta}. Further, $\langle
F^2_{\rm Na} \rangle_i$ is the sodium form factor averaged over the bin
width and $f_{\rm Na} = m_{\rm Na} / (m_{\rm Na} + m_{\rm I})$ is the
sodium mass fraction of the NaI crystal. For the modulation amplitude in
CoGeNT we proceed analogously.
Note that the conversion factor from $\bar \eta$ to $\tilde\eta$ is
the same as for $A_\eta$ to $\tilde A_\eta$, and does not dependent on
the nucleus. Therefore, the bounds \eqref{eq:bound_gen2} and
\eqref{eq:bound_spec2} apply to $\tilde \eta$, $\tilde A_\eta$ without
change, even if the l.h.s. and r.h.s. refer to different experiments. 

Let us briefly describe the data we use to derive the upper bounds on
$\tilde\eta$. We consider results from XENON10~\cite{Angle:2011th}
(XE10) and XENON100~\cite{Aprile:2011hi} (XE100). In both cases we
take into account the energy resolution due to Poisson fluctuations of
single electrons. XE100 is sensitive to the interesting region of $v_m$
only because of upward fluctuations from below the threshold. We adopt
the best-fit light-yield efficiency $L_{\rm eff}$ from
\cite{Aprile:2011hi}. The XE10 analysis is based on the so-called S2
ionization signal which allows to go to a rather low threshold. We
follow \cite{Angle:2011th} and impose a sharp cut-off of the
efficiency below the threshold.  From CDMS we use results from a
dedicated low-threshold (LT) analysis~\cite{Ahmed:2010wy} of Ge data,
as well as data on Si \cite{Akerib:2005kh}.  In the case of SD
scattering on protons particularly strong bounds are obtained from
experiments with a fluorine target. We consider the results from
SIMPLE \cite{Felizardo:2011uw}, which uses F$_5$C$_2$Cl. We use the
observed number of events and expected background events to calculate
the combined Poisson probability for Stage 1 and 2. For the prediction
we include energy dependent threshold efficiencies from
\cite{Felizardo:2011uw}.

For all experiments we use the lower bound on the expected events,
Eq.~\eqref{eq:mu}, to calculate the probability of obtaining 
less or equal events than observed. For XE100, CDMS Si, and SIMPLE we just use
the total number of events in the entire reported energy range. For
XE10 and CDMS LT the limit can be improved if data are binned and the
corresponding probabilities for each bin are multiplied. This assumes
that the bins are statistically independent, which requires to make
bins larger than the energy resolution. For XE10 we only use two
bins. For CDMS LT we combine the 36 bins from Fig.~1 of
\cite{Ahmed:2010wy} into 9 bins of 2~keV where the energy resolution
is 0.2~keV.

\begin{figure}
  \includegraphics[width=0.40\textwidth]{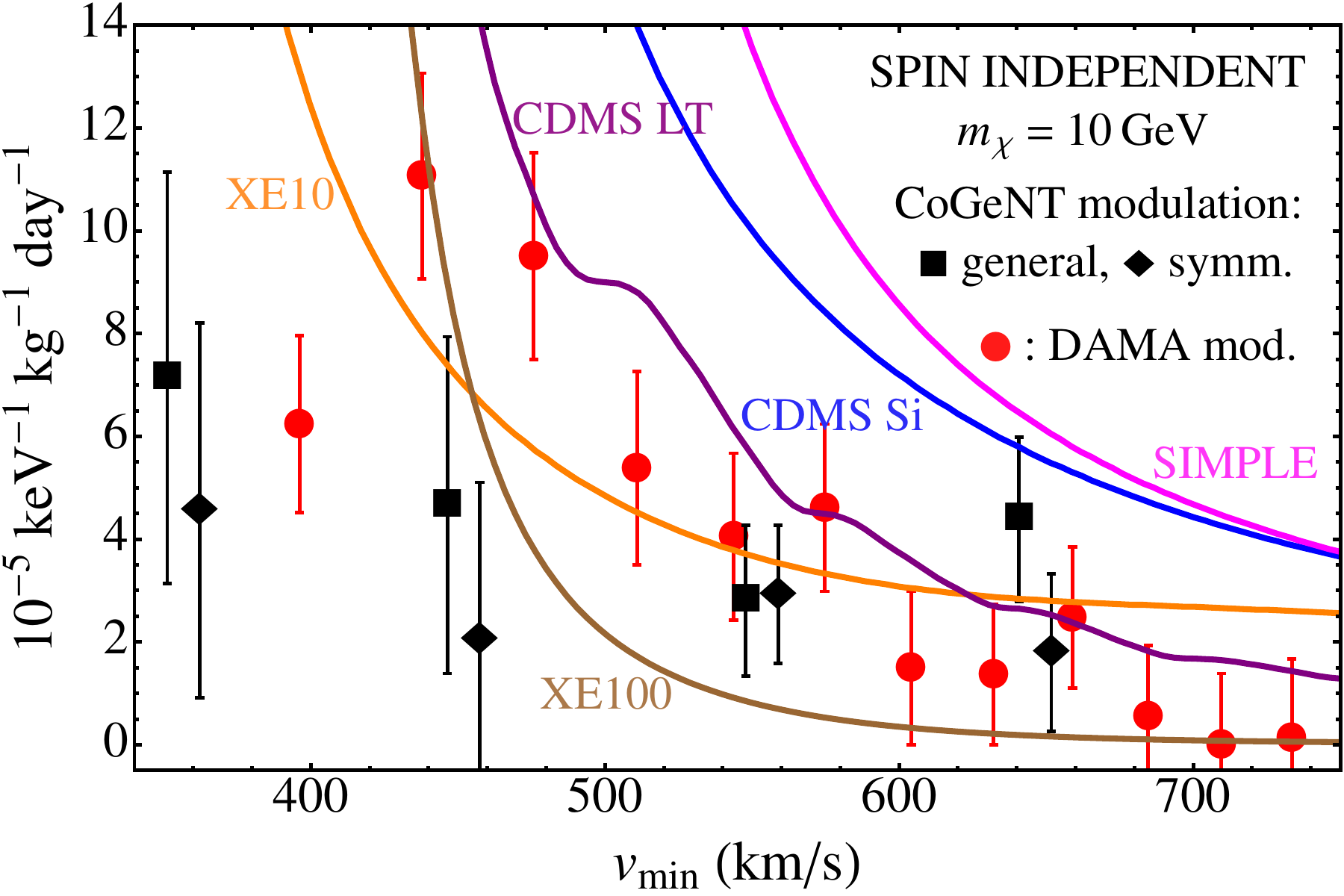}
  \caption{\label{fig:rates} Upper bounds on $\tilde \eta$ at $3\sigma$ from
  XENON100, XENON10, CDMS LT, CDMS Si, and SIMPLE. 
  The modulation amplitude $\tilde A_\eta$ is shown for DAMA (for $q_{Na}=0.3$) and
  CoGeNT for both free phase fit (general) and fixing the phase to June 2nd
  (symmetric). We assume a DM mass of 10~GeV and SI interactions.}
\end{figure}

\begin{figure}
  \includegraphics[width=0.4\textwidth]{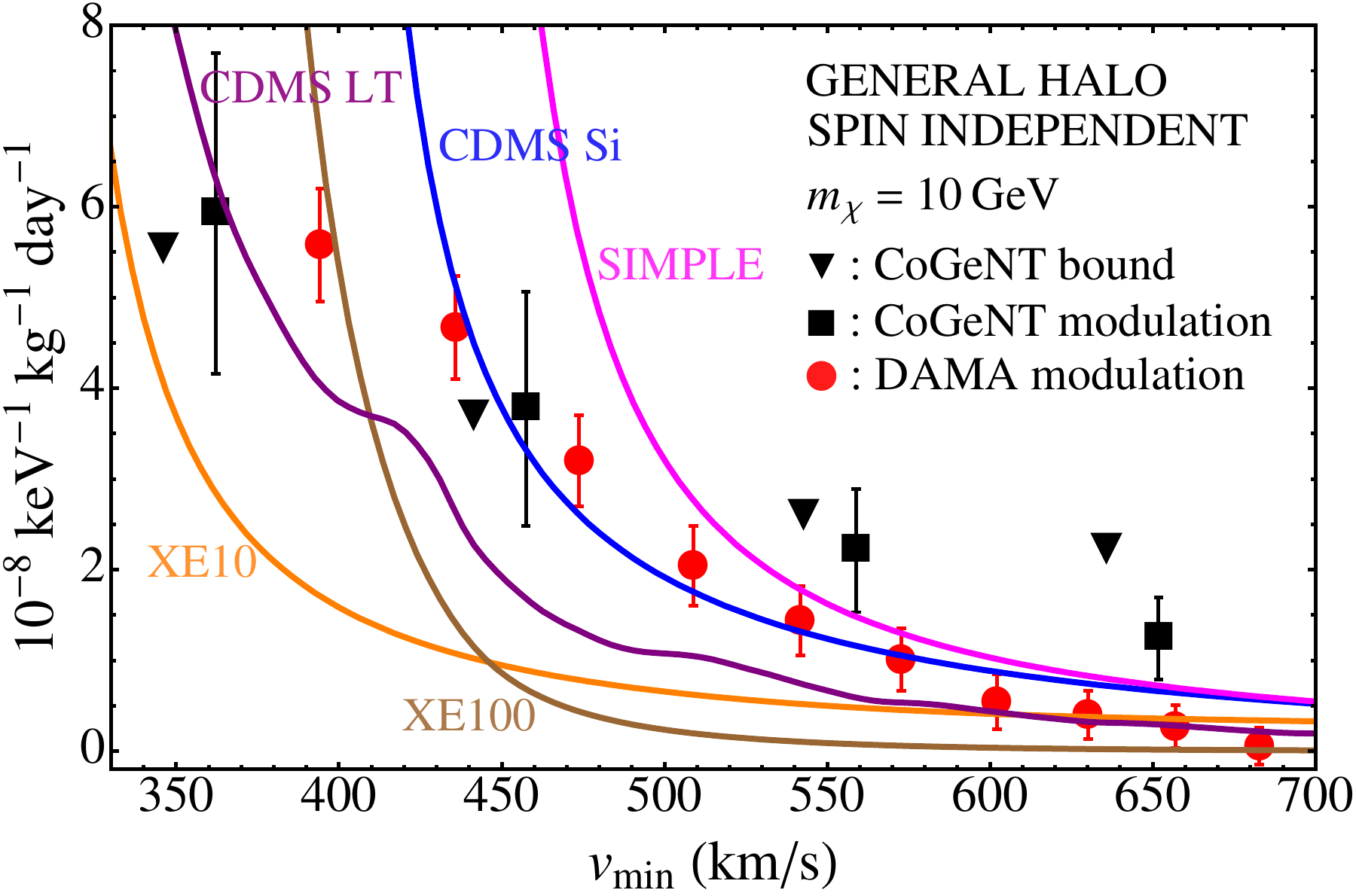}
  \caption{\label{fig:SI} Integrated modulation signals, $\int_{v_1}^{v_2} d v A_{\tilde \eta}$, from DAMA and
    CoGeNT compared to the $3\sigma$ upper bounds for the general
    halo, Eq.~\eqref{eq:bound_gen2}. We assume SI interactions and a
    DM mass of 10~GeV. The integral runs from $v_1=v_{\rm min}$ till 
    $v_2=743$ km/s (end of the 12th bin in DAMA).}
\end{figure}

{\bf Results.} In Fig.~\ref{fig:rates} we show the 3$\sigma$ limits
(CL~$= 99.73\%$) on $\tilde \eta$ compared to the modulation
amplitudes $\tilde A_\eta$ from DAMA and CoGeNT for a DM mass of
10~GeV. Similar results have been presented in \cite{Frandsen:2011gi,
  Gondolo:2012rs}.  The CoGeNT amplitude depends on whether the phase
is floated in the fit or fixed at June 2nd \cite{Fox:2011px}, which
applies to the ``general'' and ``symmetric'' halos, respectively.
Already at this level XE100 is in tension with the modulation from
DAMA (and to some extent also CoGeNT). 

We now apply our method. As shown in Fig.~\ref{fig:SI} the null search results become significantly more constraining after applying the bounds on the integrated annual modulation $\int_{v_1}^{v_2}dv A_{\tilde \eta}$
from Eq.~\eqref{eq:bound_gen2}. DAMA
and GoGeNT are strongly excluded by the bounds from XE100, XE10, CDMS
LT even for the general halo. If one were to assume in addition that  the halo is
symmetric, the bounds would get even stronger. Then also CDMS Si excludes DAMA,
and there is some tension with SIMPLE (not shown).

\begin{figure}
 \includegraphics[width=0.4\textwidth]{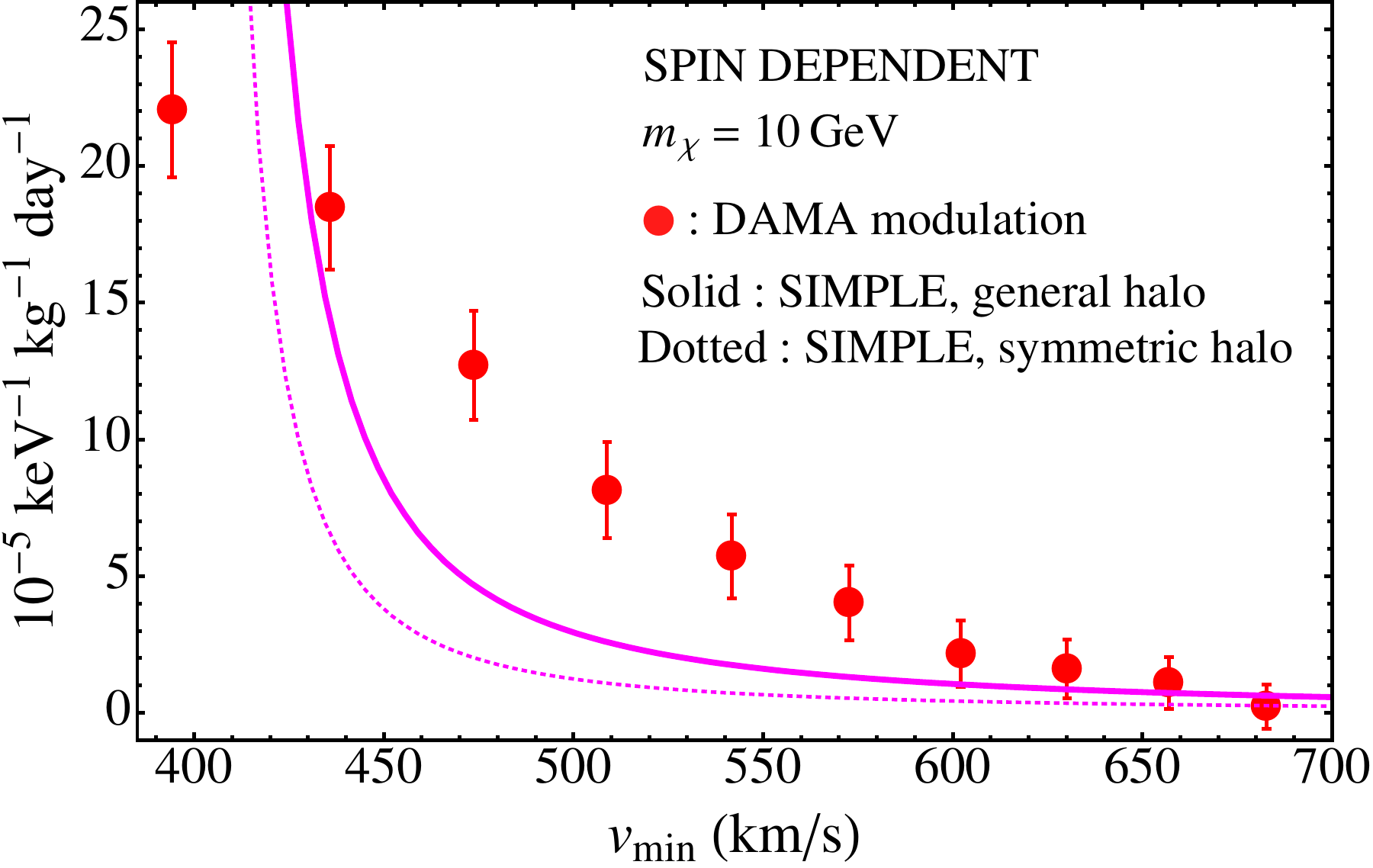}
  \includegraphics[width=0.39\textwidth]{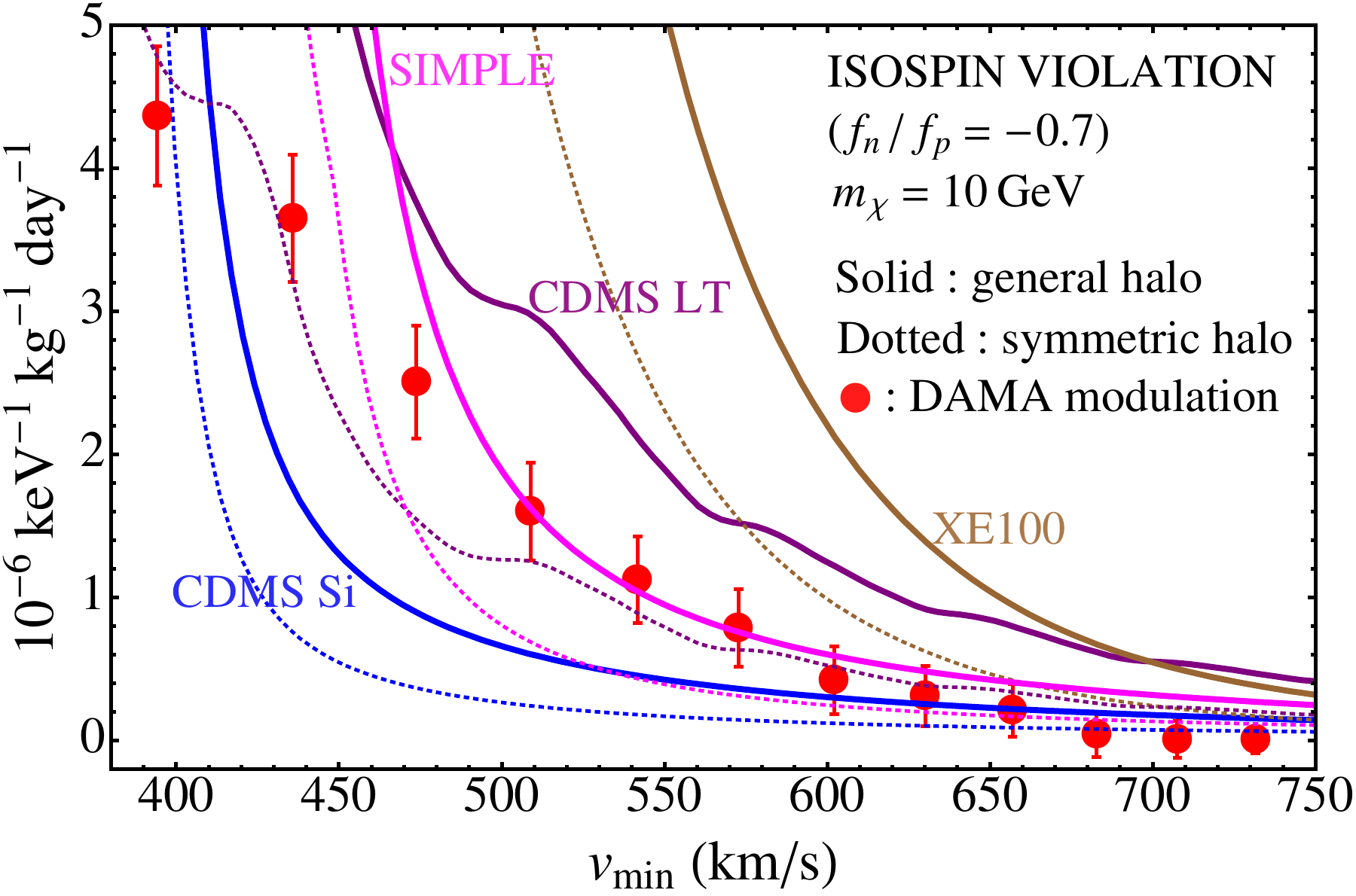}
 \caption{\label{fig:SD-IV} Integrated modulation signal 
   $\int_{v_{\rm min}}^{v_2}dv A_{\tilde \eta}$ from DAMA
   compared to the $3\sigma$ upper bounds for the general halo,
   Eq.~\eqref{eq:bound_gen2} (solid), and symmetric halo,
   Eq.~\eqref{eq:bound_spec2} with $\sin\alpha = 0.5$ (dotted). We
   assume a DM mass of 10~GeV, and SD interactions on protons (upper
   panel) and SI interactions with $f_n/f_p = -0.7$ (lower panel). 
   The upper limit of the integration is $v_2=743$~km/s.}
\end{figure}

In Fig.~\ref{fig:SD-IV} we consider two variations of
DM--nucleus interaction. The upper panel is for the case when
the DM particle couples to the spin of the proton. The null search result of
Xe and Ge experiments are then irrelevant. However, the bound from SIMPLE
is in strong disagreement with the modulation signal in DAMA, due to
the presence of fluorine in their target. (A comparable limit from fluorine
has been published recently by PICASSO~\cite{Archambault:2012pm}.) In
the lower panel of Fig.~\ref{fig:SD-IV} we show the case of SI isospin violating interactions
with $f_n/f_p=-0.7$. This choice evades bounds from Xe, but now the
DAMA modulation is excluded by the bounds from CDMS~Si for the general
halo and CDMS Si, LT, and SIMPLE for the symmetric halo.

Let us now quantify the disagreement between the observed DAMA
modulation and the rate from another null-result experiment using our
bounds. We first fix $v_m$. To each value of $\tilde\eta(v_m)$
Eq.~\eqref{eq:mu} provides a Poisson mean $\mu(v_m)$. We can then
calculate the probability $p_\eta$ to obtain equal or less events than
measured by the null-result experiment. Then we construct the bound on
the modulation using the same value $\tilde \eta(v_m)$ on the
r.h.s.\ of Eq.~\eqref{eq:bound_gen2} or \eqref{eq:bound_spec2} (the
integrand $\tilde\eta(v)$ in Eq.~\eqref{eq:bound_gen2} is calculated
using the same $p_\eta$ but with $v>v_m$ in Eq.~\eqref{eq:mu}). We
calculate the probability $p_A$ that the bound is not violated by
assuming on the l.h.s.\ of Eq.~\eqref{eq:bound_gen2} or
\eqref{eq:bound_spec2} a Gaussian distribution for the DAMA modulation
signal with the measured standard deviations in each bin. Then $p_{\rm
  joint}(\tilde\eta) = p_\eta p_A$ is the combined probability of
obtaining the experimental result for the chosen value of
$\tilde\eta$. Then we maximize $p_{\rm joint}(\tilde\eta)$ with
respect to $\tilde\eta$ to obtain the highest possible joint
probability.

The results of such an analysis are shown in Fig.~\ref{fig:prob}. The
analysis is performed at the fixed $v_m$ corresponding to the 3rd
modulation data point in DAMA, depending on the DM mass $m_\chi$. We
find that for all considered interaction types and $m_\chi \lesssim
15$~GeV at least one experiment disfavors a DM interpretation of the
DAMA modulation at more than $4\sigma$ even under the very modest
assumptions of the ``general halo''. In the case of SI interactions
the tension with XE100 is at more than $6\sigma$ for $m_\chi \gtrsim
8$~GeV and saturates at the significance of the modulation data point
itself at about $6.4\sigma$ for $m_\chi \gtrsim 13$~GeV. The exclusion
from XE10 is nearly independent of the DM mass slightly below
$6\sigma$.  We show also a few examples of the joint probability 
in case of a ``symmetric halo'' (dashed curves).

\begin{figure}[t]
  \includegraphics[width=0.4\textwidth]{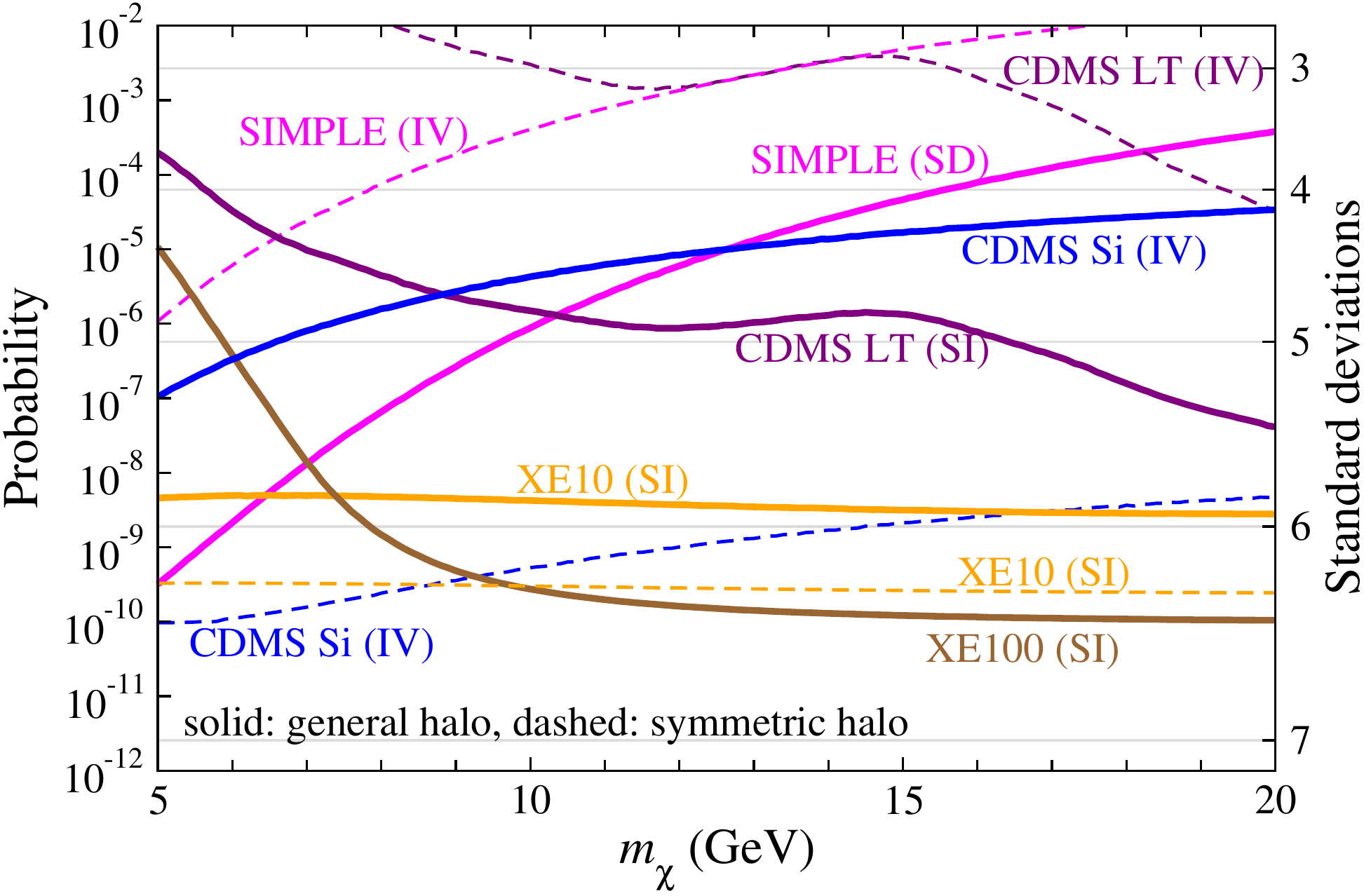}
  \caption{\label{fig:prob} The probability that the integrated modulation
  amplitude in DAMA (summed starting from the 3rd bin) is compatible with
  the bound derived from the constraints on $\tilde\eta$ for various
  experiments as a function of the DM mass. The label SI (SD), refers to
  spin-independent (spin-dependent) interactions with $f_n=f_p$ ($f_n = 0$),
  and IV refers to isospin-violating SI interactions with $f_n/f_p = -0.7$.
  For solid and dashed curves we use the bounds from
  Eqs.~\eqref{eq:bound_gen2} and \eqref{eq:bound_spec2}, respectively.}
\end{figure}

While astrophysics uncertainties are avoided, the obtained bounds are
still subject to nuclear, particle physics and experimental uncertainties. For instance, the tension
between the DAMA signal and the bounds depends on the value of the Na
quenching factor $q_{\rm Na}$, light yield or ionization yield
efficiencies in Xe, upward fluctuations from below threshold, and so
on. For example, if a value of $q_{Na} = 0.45$ is adopted instead of
the fiducial value of 0.3 consistency for SD and isospin violating
interactions can be achieved in the case of the general halo at around
$3\sigma$, while for SI interactions the XE10 bound still implies
tension at more than $5\sigma$ for $m_\chi \gtrsim 10$~GeV. Hence, the
precise CL of exclusion may depend on systematic uncertainties.

In conclusion, we have presented a powerful method to check the
consistency of an annual modulation signal in a DM direct detection
experiment with bounds on the total DM scattering rate from other
experiments, almost completely independent of astrophysics, for a
given type of DM--nucleus interaction. While our bounds strongly
disfavor a DM interpretation of present annually modulated signals in
the case of SI and SD elastic scattering, the method will be an
important test that any future modulated signal will have to pass
before a DM interpretation can be accepted.

{\bf Acknowledgements:} 
J.H.-G.\ is supported by the MICINN under the FPU program.


\end{document}